# a Tale of Two Turing Machines

Eric C.R. Hehner

Department of Computer Science, University of Toronto
hehner@cs.utoronto.ca

**Abstract:** Two Turing Machines may be able to answer questions about each other that they cannot answer about themselves.

## Red Machine

In a corner of a room, there is a Turing Machine [0], or TM. The wall beside where it sits is red, and for that reason, the TM is affectionately known as "Red", although it is not actually red. Being old, it has no electronic communication ability. For input, a human has to do something (maybe push buttons); for output, a human must look at its paper tape. It has an infinitely long tape, fully loaded with every TM program. Each program is a finite sequence of TM instructions, and there are infinitely many programs, and they are all on the tape.

Is one of the programs a function that, given input $p$, computes the answer to the question "Does execution of program $p$ terminate?"? Let's suppose one of the programs is such a function, and let's call it *halts*. Since all programs are there, there is a program, let's call it *diag*, whose execution is as follows:
      *diag* calls *halts* to determine if its own (*diag*'s) execution will terminate;
      if *halts* reports that *diag*'s execution will terminate,
          then *diag*'s execution becomes a nonterminating loop,
          otherwise *diag*'s execution terminates.
So whatever *halts* reports, it is wrong. And we have our answer: none of the programs on Red's tape determines halting for all programs on the tape. As far as I know, this result is universally accepted.

It is almost universally accepted that the reason there is no *halts* program is that a TM is not computationally powerful enough to perform the task; that's the definition of "incomputable". But I believe that the reason there is no *halts* program is that the task is self-contradictory, or inconsistent. Nothing can perform a self-contradictory task, no matter how powerful it is.

## Blue Machine

In the opposite corner of the same room there is another TM. The wall beside where it sits is blue, and for that reason, this TM is affectionately known as "Blue", although it is not actually blue. The two machines are actually the same color. In fact, the two machines are identical in every respect except identity: they are identically built, and their tapes have identical contents, but they are in different locations, and they have different names. However, where a machine sits, and what its name is, do not in any way affect its operation: the two machines behave identically.

For exactly the same reason that there is no *halts* program on Red, there is also no *halts* program on Blue: none of the programs on Blue's tape determines halting for all programs on Blue's tape, because that task is self-contradictory.

# Red and Blue Machines Thinking about Each Other

Is there a program on Red that answers the question "Am I the Red machine?"? Of course there is: its execution just prints "yes". And since the Red and Blue machines are identical, the very same program exists on Blue: it also prints "yes". But this program on Blue does not answer the question "Am I the Red machine?". Although the programs are identical, they do not answer the same question. But there is a program on Blue that answers the question "Am I the Red machine?". It just prints "no". To answer the same question, we need a different program, due to the self-reference.

Is there a program on Red that answers the question "Can the Red machine correctly answer "no" to this question?"? On Red, there is a program that prints "yes", but that answer says that "no" is the correct answer. There is another program that prints "no", but that answer says that Red cannot do what it is doing (printing "no" in answer to the question). There is no program on Red that answers the question correctly. That is not because the task requires more computing power than a TM can offer, and is therefore "incomputable"; it is because the task is self-contradictory. But there is a program on Blue that answers that same question correctly: it prints "no". Blue correctly says that Red cannot correctly answer "no" to the question. Due to the twisted self-reference, the task was impossible for Red, but possible for Blue.

Symmetrically, there is no program on Blue that can answer the question "Can the Blue machine correctly answer "no" to this question?". But Red can answer it: "no". Obviously, we cannot conclude that each of these identical TMs is more powerful than the other.

## How to Compute Halting

In some ways, the halting problem is like the problem in the previous two paragraphs. On Red, the *halts* program must report the halting status of programs that call *halts*, thus creating a self-reference. By placing a twist in that self-reference loop, it is a self-contradictory, impossible task for Red to perform. But maybe there is a program on Blue's tape that, given input *p*, computes the answer to the question "Does execution of program *p* on Red's tape terminate?". Let me suppose there is, and call it *Redhalts*. Programs on one machine have no way to call programs on the other machine, so there is no program on Red that calls *Redhalts* and then does the opposite.

There is a program on Red that is identical to *Redhalts*; let me call it *Redhalts*. But we know that there is no program on Red to determine halting on Red. The resolution of the apparent inconsistency comes from the preceding section: identical programs on the two machines do not necessarily answer the same question. It is possible that *Redhalts*, residing on Blue, determines halting on Red, but the identical program *Redhalts*, residing on Red, does not.

## Conclusion

The usual textbook proof that halting is incomputable does not prove that halting is incomputable. It proves that the specification "Write a program in a TM-equivalent language to determine whether programs in that same language halt." is inconsistent. It may be possible to write a program in TM-equivalent language A to determine if programs in TM-equivalent language B halt, if B programs cannot call A programs. If so, halting is computable.

**Addendum** added 2016-10-25 in reply to a question

I was asked "How does the computer know what question to answer?".

Does a computer "think" or "know" anything? The question does not call for experimentation or observation; it's simply a linguistic question. We have collectively decided to say that airplanes fly, like birds do, even though airplanes are not alive. We have collectively decided not to say submarines swim. I know what computers do from the level of atoms right up to the level of programs. But that doesn't tell me what to call that activity. We, the English speakers of the world, have apparently decided to say that computers think and know, and I have decided to go along with that decision.

I can write a computer program that answers the question "What is the first letter of the Roman alphabet?". It just prints the letter "A". Whenever I want to know the first letter of the alphabet, I can run this program, and it tells me. How do I know to run this particular program? In a modern programming language, I might give this program the name *firstletter*, which is mnemonic. But TM programs don't have names. So I, a human, keep a piece of paper on which I have written various questions, and for each, the address of the program to run to answer it. Another of my questions is "What letter of the Roman alphabet comes before "B"?". And my piece of paper tells me the same address as before. Whenever I run the program that just prints "A", you might say, if you are inclined to use language this way, that the computer knows that I have asked a question whose answer is "A", but not which one.

On my piece of paper, I also have the question "Are you Red?", and the address of the program to invoke on Red. When I invoke this program on Red, I get the correct answer: "yes". Whenever I run the program that just prints "yes", you might say, if you are inclined to use language this way, that the computer knows that I have asked one of the questions whose answer is "yes", but not which one. Although Red does not know exactly what question it is answering, it answers anyway. When I invoke the same program on Blue, which is identical to Red, I get an incorrect answer. To get the correct answer on Blue, I have to invoke a different program.

Suppose there is a *halts* program that works on all programs except those that (directly or indirectly) call *halts*. I invoke it, and feed in program *gcc*. The computer knows which program I am asking about because I told it: *gcc*. But does the computer know that it is determining the halting status of *gcc*? The name *halts* is a clue to me, but not to the computer, and I could have called the program *fred*.

Should the *halts* program start with the question "What is the halting status of the given program?". Well, yes, but as a comment for humans. It doesn't help the computer to know anything. It's still the halting program even without that comment. And TM programs don't have comments.

I know that right now I am thinking about computers. To know that, I not only need to be thinking about computers, I also need to have, and use, a self-reflective ability. I need to think: "I am thinking about computers.". If we haven't invoked a self-reflective program in the computer, then I would say no, the computer does not know it is computing the halting status of *gcc*, but it computes the halting status of *gcc* anyway.

[other papers on halting](other papers on halting)